\documentstyle[aps,prl,multicol,epsf]{revtex}  
\begin{document}    
\draft  
\title{Shot noise in chaotic systems: ``classical'' to quantum crossover}      
\author{Oded Agam$^1$, Igor Aleiner$^{2,3}$, and   
Anatoly Larkin$^{4,5}$}     
\address{$^1$The Racah Institute of Physics, The Hebrew University,   
Jerusalem 91904, Israel\\  
$^2$Physics and Astronomy Department, SUNY Stony Brook, Stony Brook,  
NY 11794\\  
$^3$Center for Advanced Studies, Drammensveien 78, N-0271 Oslo, Norway\\  
$^4$Theoretical Physics Institute, University of Minnesota, Minneapolis,  
MN 55455\\  
$^5$L.D. Landau Institute for Theoretical Physics, 117940 Moscow, Russia  
}  
\maketitle  
\begin{abstract}  
This paper is devoted to study of the classical-to-quantum crossover  
of the shot noise in chaotic systems. This crossover is  
determined by the ratio of the particle dwell time in the system,  
$\tau_d$, to the characteristic time for diffraction $t_E \simeq  
\lambda^{-1} |\ln \hbar|$, where $\lambda$ is the Lyapunov exponent.  
The shot noise vanishes when $t_E \gg \tau_d $, while reaches  
a universal value in the opposite limit. Thus, the Lyapunov  
exponent of chaotic mesoscopic systems may be found by  shot noise  
measurements.  
\end{abstract}       
\begin{multicols}{2}  
  
The shot noise in the current flowing through a system  
is a random process associated with the nonequilibrium state  
into which the system is driven by the applied voltage\cite{review}.   
The zero-frequency power spectrum of the noise is given by  
\begin{eqnarray}  
S= 2 \int_{-\infty}^{\infty} dt'  
\left[\langle I(t) I(t') \rangle -\langle I\rangle^2\right]=  
2Fe\langle I\rangle,  
\label{cccf}  
\end{eqnarray}   
where $-e$ is the charge of the electron, and  
$\langle I\rangle$ is the average current flowing through the system.  
In vacuum tubes, the shot noise emerges from the Poisson  
distribution of the transmission of uncorrelated electrons.     
The result, in this case, is given by Schottky\cite{Schottky18}  
formula (\ref{cccf}), with $F=1$.   
However, in more general cases, the charge carriers are correlated,   
for example due to Fermi statistics. These correlations suppress the noise,   
compared to Schottky result, by a factor known as the Fano factor, $F$.   
In diffusive wires with noninteracting electrons\cite{Beenakker92,1/3}   
$F=1/3$, while in quantum dots\cite{1/4}, $F=1/4$.   
These suppression factors are universal in the sense that they   
are independent of the details of the systems.  
  
In a view of this universality, an intriguing feature of these results  
is the absence of an explicit $\hbar$ dependence in $F$.  
This is despite the fact that the   
source of the shot noise is quantum mechanical uncertainty.
Indeed, it has been shown by Beenakker and van 
Houten\cite{Beenakker91} that the shot noise vanishes in 
the classical limit. The purpose of this Letter is to characterize the  
quantum-classical crossover of the shot noise in noninteracting  
mesoscopic systems. It will be shown this crossover is governed by the  
ratio of two time scales: the average dwell time, $\tau_d$, which is  
the typical time the electron stays in the system, and the Ehrenfest  
time, $t_E\propto |\log \hbar| $ which is the time scale 
where quantum effects set in.
 
A qualitative understanding of the quantum-classical crossover of the  
shot noise for Fermions can be achieved by examining the general  
formula\cite{Lesovik89,Beenakker92} at zero frequency   
{\narrowtext     
\begin{figure}[ht]       
  \begin{center}      
\leavevmode      
        \epsfxsize=8cm         
         \epsfbox{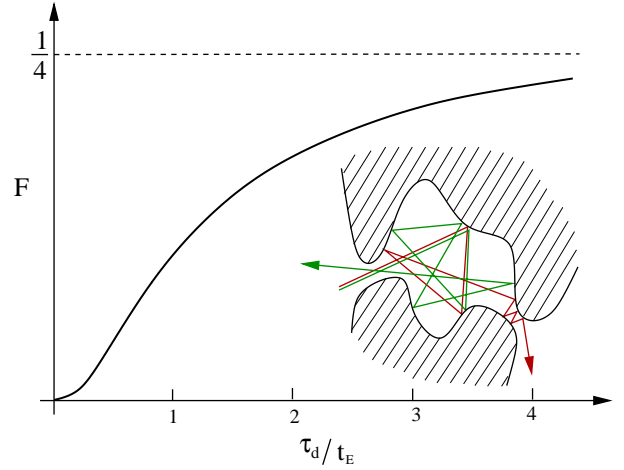}       
\end{center}        
	\caption{The classical-to-quantum crossover of the   
shot noise in chaotic dots. The solid line shows the Fano factor   
(\protect\ref{fff})  
as  function of the ratio between the dwell time of   
the particle in the dot, $\tau_d$, and the Ehrenfest time  
(\protect\ref{tE}).  
}                
\end{figure}         
}    
\noindent  
and temperature: $S=\mbox{Tr} \left\{ {\cal T}(1-{\cal T})\right\}  
e^3 V/2\pi\hbar$. Here $V$ is the applied voltage, and ${\cal T}$ is  
the transmission matrix. For a single channel, the factor ${\cal  
T}(1-\cal{T})$ represents the variance of the probability of a  
particle to pass through the channel.  It vanishes both when  
this probability is unity $({\cal T}=1)$ or when it is zero $({\cal  
T}=0)$. Thus, the shot noise of open or closed channels is zero
because the incident flux of Fermi particles, at zero temperature, 
is noiseless. The only possible  source of 
the noise is the probabilistic nature  
of the transmission and reflection, which implies ${\cal T}\neq 1,0$.
 
Classical dynamics, however, is deterministic. A trajectory 
of an incident particle can either pass through the system   
(${\cal T}=1$, if we loosely view the trajectory as a channel),   
or be reflected back (${\cal T}=0)$.  
Therefore, the shot noise vanishes in the classical limit.  
Yet, a quantum particle can switch between trajectories,
an event which we term as {\em diffraction}\cite{comment}.
Consequently, the wave packet of the particle splits into 
two components: one  passing  through the system and another 
which is reflected back (see inset of Fig.~1).  
The transmission probability is now  
different from one or zero, and the shot noise is finite.   
Thus the quantum-classical crossover   
of shot noise is determined by the probability of the particle to:  
(a) split its classical path by diffraction, and (b) that the resulting   
two trajectories diverge during the dwell time of the particle 
in the system.  
  
Consider a chaotic system with smooth potential energy characterized  
by a length scale $a$. If $a$ is larger than the particle wavelength,  
$\lambda_F$, the typical time for the trajectories to diverge is  
the time which takes for a minimal wave packet to spread over a  
distance of order $a$.  We picture the initial electron wave packet   
as a collection of trajectories separated by a distance ${\lambda_F}$   
[see Ref.~\cite{Aleiner96} for similar  
estimates for weak localization].  The separation among these  
trajectories after time $t$ is dominated by the unstable nature of the  
classical dynamics.  Namely, if $\rho(0)$ is the initial  
distance between a pair of trajectories, then after time $t$ the  
distance is of order $\rho(t) \simeq \rho(0) e^{ \lambda t}$, where  
$\lambda$ is the Lyapunov exponent of the system.  Thus, setting  
$\rho(t_E)=a$ and $\rho(0)={\lambda_F}$ one obtains:  
\begin{equation}  
t_E= \frac{1}{\lambda} \ln \left( \frac{a}{\lambda_F} \right)  
\label{tE}  
\end{equation}  
  
Notice that the Ehrenfest time, $t_E \propto |\ln \hbar|$,  
diverges logarithmically when $\hbar \to 0$, while  
the average dwell time of the particle, $\tau_d$, is essentially independent  
of $\hbar$. Thus, in the classical limit, the Ehrenfest time is always   
larger than the average 
dwell time, $t_E \gg \tau_d$, and the shot noise is zero.   
The universal results for the shot noise mentioned above, implicitly,   
assume the system to be in the opposite limit, $\tau_d \gg t_E$.   
In this regime the particle stays in the sample long enough to experience   
diffraction, and as a result the shot noise is finite, see inset of Fig.~1.   
  
So far, our discussion applies to any chaotic system.  
To make it more specific, consider the  
shot noise of quantum dots having a non-integrable shape. Let us assume  
the elastic relaxation time within the dot to be much shorter than the  
Ehrenfest and  dwell times, and the contacts with the leads  
to have an equal large number of channels. Let the leads be in thermal  
equilibrium, and the bias between them be $V$.  
As we will show the zero-frequency correlator, at zero temperature, 
is given by Eq.~(\ref{cccf}) with the Fano factor:
\begin{equation}  
F = \frac{1}{4}\Gamma;\quad \Gamma=\exp \left[-\frac{ t_E}{\tau_d}\left(1-   
\frac{\lambda_2}{2\lambda^2\tau_d}  
+\dots\right)  \right],  
\label{fff}  
\end{equation}  
where $\lambda_2$ is of order $\lambda$. 
The behavior of this Fano factor is depicted in Fig.~1.  
  
Before turning to the rigorous derivation of Eq.~(\ref{fff}), we
discuss the physical meaning of this result.  Factor $F$ consists of
the quantum value, $ 1/4$, multiplied by the probability of the
particle to diffract during the dwell time, see also Ref.~
\cite{Aleiner96}.  Indeed, for the chaotic dots the distribution of
the dwell time, $t$, is Poissonian, $P(t) = e^{- t/\tau_d}/\tau_d$,
where $\tau_d$ is the average escape time. As we already explained,
only the trajectories with dwell time lareger than $t_E$ contribute to
the noise. Thus, the relative number of such trajestories is $e^{-
t_E/\tau_d}$ which is the first term in the exponent of factor
$\Gamma$ in Eq.~(\ref{fff}). The second factor in the exponent take
into account the deviations in the value of the Lyapunov exponent
$\delta\lambda$ for different trajectories, $\langle\delta
\lambda^2\rangle =\lambda_2/t$.

We turn, now, to the formal derivation of the above results.  We  
choose to work in units where $\hbar=k_B=1$. The dependence   
on these constants will be restored in  
our final expressions. The derivation\cite{Nazarov94} is based on the  
Keldysh formalism\cite{Keldysh64}.  
      
The matrix Green function of our system  
\begin{equation}  
{\cal G}(\zeta;\zeta')=   
\pmatrix{ {\cal G}^R(\zeta;\zeta') &  
{\cal G}^K(\zeta;\zeta') \cr  
{\cal G}^Z(\zeta;\zeta')  &  {\cal G}^A(\zeta;\zeta')  
}  
;\quad \zeta=({\bf r},t)  
\end{equation}  
($R,A,K$ denote retarded, advanced, and Keldysh  
components respectively) satisfies the Schr\"{o}dinger equation:  
\begin{equation}  
\left[\left(i \frac{\partial}{\partial t} -\hat{H}  
\right)   
\hat{\bf 1}  
- \hat{\sigma}_x \hat{\mbox{\boldmath $A$}} \cdot \hat{\mbox{\boldmath $J$}}  
 \right] {\cal G} = \hat{\bf 1}\delta(\zeta-\zeta'),  
\label{Sheeq}  
\end{equation}  
where $\hat{H}$ is the   
Hamiltonian of the system,  
$\sigma_x\hat{\mbox{\boldmath $A$}}(\zeta) \cdot \hat{\mbox{\boldmath  
$J$}}$ is a source term,   
$\sigma_x$ is the   
first Pauli matrix in Keldysh space, and $\hat{\mbox{\boldmath  
$J$}}$ is the current density operator:  
\[  
\hat{\mbox{\boldmath $J$}}  f({\bf r},{\bf r}')=    
-\left. \frac{i e}{2 m} \left(  
\frac{ \partial }{\partial {\bf r}} -   
\frac{ \partial }{\partial {\bf r}'} \right)  
f({\bf r},{\bf r}')\right|_{ {\bf r} \to {\bf r}'}.  
\]  
The variational  
derivative  
\begin{equation}  
\left\langle \left\{\hat{j}_\alpha(\zeta); \hat{j}_\beta (\zeta')\right\}   
\right\rangle =  
\frac{1}{2} \mbox{Tr} \left\{ \sigma_x \hat{J}_\alpha \frac{ \delta   
{\cal G}(\tilde{\zeta};\zeta)}  
{ \delta A_\beta (\zeta')} \right\}_{\tilde{\zeta} \to \zeta,   
{\bf A}=0} \label{dccf}  
\end{equation}  
gives the current density correlation function.  
Hereinafter,   
$\left\{\hat{A};\hat{B}\right\}\equiv\hat{A}\hat{B}+\hat{B}\hat{A}  
$, for any operators $\hat{A}$ and $\hat{B}$.   
Integrating Eq.~(\ref{dccf}) over the cross sections of the leads  
yields the total current correlation function (\ref{cccf}).  
  
Our main purpose is to reduce the above equations to simpler ones
which hold in the semiclassical limit, $a \gg \lambda_F$.  However,
since the transition bewtween classical trajectories plays a
crucial role here, it is difficult to implement periodic orbit
theory\cite{Gutzwiller90} for this purpose, see e.g. \cite{Whitney98}. The
standard diagrammatic technique\cite{AGD} is suitable neither, because
it is technically difficult to take the classical correlations into
account.  To circumvent this difficulty, we follow
Ref.~\cite{Aleiner96} and add a weak random potential, $V({\bf r})$,
to the semicalassical potential $U({\bf r})$.  The random potential,
$V({\bf r})$, generates a small angle scattering that models the
diffraction; the total Hamiltonian of our system is
$\hat{H}=\hat{H}_0+V({\bf r})$, where $\hat{H}_0= {\bf p}^2/2m +
U({\bf r})$ is the bare Hamiltonian of the system. To mimic the
diffraction effects, the strength of $V({\bf r})$ is chosen such that
the transport mean free time for the scattering on $V({\bf r})$ is
given by 
\begin{equation}
1/ \tau_{tr} = \lambda\lambda_F/a.
\label{tautr}
\end{equation}
The numerical coefficient in Eq.~(\ref{tautr}) is not important for
the calculation with logarthmical accuracy; value of $\tau_{tr}$
enters the final result only througth $t_E=(1/\lambda)\ln (\lambda
\tau_{tr})$,
see Refs.~\cite{Aleiner96,pre} for further discussion.
  
The construction of the semiclassical approximation of the Green  
function follows the usual steps\cite{Keldysh64,Eilenberger}.  First,  
the operators are Wigner transformed:  
\[  
{\cal G}_W({\bf R},{\bf P};t,t')= \int d {\bf r}e^{i {\bf P}\cdot {\bf r}}  
{\cal G}\left({\bf R}- \frac{\bf r}{2}, t;   
{\bf R}+\frac{\bf r}{2},t' \right).  
\]  
Next, ${\cal G}$ is averaged over the disorder in the self-consistent  
Born approximation, and the Green function is projected on the energy  
shell. The projection is obtained by changing the coordinates from  
$({\bf R}, {\bf P})$ to $(\xi,{\bf x})$ where $\xi ={\cal H}_0({\bf  
R}, {\bf P})- \epsilon_F$ is the distance along the energy axis in  
phase space, and ${\bf x}=({\bf R},{\bf n})$ are coordinate on the  
energy shell, ${\bf n}$ being a unit vector in the momentum direction. 
The outcome of the procedure is that the on-shell matrix  
Green function,
\begin{equation}  
g({\bf x};t,t') = \frac{i}{\pi\nu} \int  \! d \xi    
~{\cal G}_W(\xi,{\bf x};t,t'),  
\label{g}  
\end{equation}  
where $\nu$ is the Weyl density of states, satisfies the   
Boltzmann-like equation:  
\begin{equation}  
\left[ \frac{\partial}{\partial t} +\frac{\partial}{\partial t'}   
+ \hat{\cal L} + \hat{O} \right] g({\bf x};t',t') =  I[g]. \label{mainEq}  
\end{equation}  
Here,  the Liouville operator is given by  
\begin{equation}  
\hat{\cal L} = \frac{\partial H_0}{\partial {\bf P}}
 \frac{\partial}{\partial {\bf R}}  
-\frac{\partial H_0}{\partial {\bf R}} \frac{\partial}{\partial {\bf P}},    
\end{equation}  
while the source operator $\hat{O}$ is   
\[  
\hat{O} g= i e {\bf v}_F\left[\mbox{\boldmath  
$A$}({\bf x}, t')   
g\hat{\sigma}_x - \mbox{\boldmath $A$}({\bf x},t) \hat{\sigma}_x g  
\right],   
\quad  
{\bf v}_F={ v}_F\mbox{\boldmath $n$},  
\]  
and $v_F$ is the Fermi velocity of the electrons.  
Finally, the probabilistic effects in the problem are introduced by   
\[  
I[g]= -\frac{1}{2 \tau_{tr}} \left[\mbox{\boldmath $\nabla$}^2_{\bf n}  
 g  * g -g *\nabla^2_{\bf n} g  \right],  
\quad  
\mbox{\boldmath $\nabla$}_{\bf n} = \mbox{\boldmath $n$}   
\times \frac{\partial}{ \partial\mbox{\boldmath $n$}}  
\]  
where $\tau_{tr}$ is given by (\ref{tautr}), 
and the convolution of two functions, say $f$ and $g$, is defined as:  
\[  
(f*g)(t,t')= \int d\tilde{t}~ f(t,\tilde{t}) g(\tilde{t},t').  
\]  
The homogeneous Eq.~(\ref{mainEq})  is supplied with the   
constraint  
\begin{equation}  
g*g=  \delta(t-t'). \label{constraint}  
\end{equation}      
  
In order to find (\ref{dccf}), Eqs.~(\ref{mainEq} -- \ref{constraint}) 
should be solved in first order in {\boldmath ${A}$}:  
$g= g_0+g_1$ where $g_0$, $g_1$ are zeroth and first order  
in {\boldmath ${A}$} respectively.  
The zeroth order term is  
\begin{equation}  
g_0=   
\pmatrix{  
\delta(t-t'); &   
\int\frac{d\epsilon}{2\pi}  
e^{i\epsilon(t-t')}  
g_0^K\left({\bf x};\frac{t+t'}{2},\epsilon\right)  
 \cr 0; & -\delta(t-t')  
},   
\end{equation}  
where the Keldysh component satisfies the equation  
\begin{mathletters}  
\label{g0K}  
\begin{equation}  
\left[ \frac{\partial}{\partial t}   
+ \hat{\cal L}- \frac{1}{\tau_{tr}}\nabla^2_{\bf n}   
\right]  g_0^K({\bf x};t,\epsilon) =0.  
\end{equation}  
The leads are assumed to be   
in thermal equilibrium, therefore, the boundary conditions are  
\begin{equation}  
g_0^K({x}\to \pm \infty;\epsilon) =2  
\tanh \left(\frac{\epsilon \pm eV/2}{2T}\right),  
\end{equation}  
\end{mathletters}  
where $T$ is the temperature.
  
Calculating the off-diagonal components of   
\[  
g_1(\epsilon,\omega;{\bf x}) =  
\pmatrix{g_1^R & g_1^K \cr  
g_1^Z & g_1^A   
},  
\]  
where $g(\epsilon,\omega)=  
\int dtd\tau  g(t_+,t_-)e^{i\epsilon\tau + i\omega t}, \   
t_\pm=t\pm\tau/2$,   
we first solve the equation for the component $ g_1^Z$:  
\begin{mathletters}  
\label{KZ}  
\begin{equation}  
\left[-i \omega + \hat{\cal L} + \frac{1}{\tau_{tr}} \nabla^2_{\bf n}   
 \right] g_1^Z = - 2 i e {\bf v}_F \mbox{\boldmath $A$}_\omega.   
\end{equation}  
Then, we find the diagonal components  from Eq.~(\ref{constraint}) as  
$2g_1^R =  - g_0^K* g_1^Z,\ 2g_1^A =  g_1^Z* g_0^K$, and substitute  
them into the equation for the component $g_1^K$. We thus obtain  
\begin{eqnarray}  
&&\left[-i \omega + \hat{\cal L} - \frac{1}{\tau_{tr}} \nabla^2_{\bf n}   
 \right] g_1^K =  2 i e {\bf v}_F \mbox{\boldmath $A$}_\omega  
\\  
&&\quad\quad- \frac{1}{2\tau_{tr}} \nabla_{\bf n}  
\left[ g_0^K({\bf x};\epsilon\!+\!\omega)  
g_0^K({\bf x};\epsilon) \nabla_{\bf n} g_1^Z({\bf x},\omega)\right].    
\nonumber  
\end{eqnarray}  
\end{mathletters}  
Solving Eqs.~(\ref{KZ}), and substituting the results   
in Eq.~(\ref{dccf}), we find  with the help of Eq.~(\ref{g}):  
\begin{equation}  
\left\langle \left\{\hat{j}_\alpha(\zeta); \hat{j}_\beta (\zeta')\right\}  
 \right\rangle_\omega  
= \frac{ \nu e^2  v_F^2}{2 }  
  \langle \langle   
 n_\alpha n'_\beta \left( F_1+ F_2\right)\rangle\rangle,  
\label{3sum}  
\end{equation}  
where the LHS represents the $\omega$ component  
of the Fourier transform with respect to $t-t'$, and   
$\langle \langle \cdots \rangle\rangle$ denotes an angular averaging with   
respect to ${\bf n}$ and ${\bf n}'$. The entries in Eq.~(\ref{3sum}) are  
\begin{eqnarray}  
F_1&=&{\cal D}_\omega ({\bf x},{\bf x}')    
Q({\bf x}') + {\cal D}_{-\omega}({\bf x}',{\bf x})   
Q({\bf x}), \label{F}\\  
F_2&=&  \int \frac{d{\bf x}_0}{\Omega_d}  
{\cal D}_\omega({\bf x},{\bf x}_0){\cal D}_{-\omega}(   
{\bf x}',{\bf x}_0)\left[- \hat{\cal L}\!+\!  
\frac{1}{\tau_{tr}}\nabla^2_{{\bf n}_0}\right]  
 Q({\bf x}_0).  
\nonumber  
\end{eqnarray}  
where ${\Omega_d}$ is the surface of unit sphere in $d$ dimensions,
${\cal D}_\omega({\bf x},{\bf x}')$ is the classical propagator of
the system,
\begin{equation}  
\left[-i \omega + \hat{\cal L} -\frac{1}{\tau_{tr}} \nabla^2_{\bf n}
\right] {\cal D}_\omega( {\bf x},{\bf x}') = \Omega_d \delta ({\bf x}-
{\bf x}'),
\label{D}  
\end{equation}  
and $Q$ characterizes the available energy space for the  
electron-hole pairs: 
\begin{equation}  
Q(\omega;{\bf x})=   
\int d\epsilon \left[1-  
\frac{1}{4} g_0^K({\bf x};\epsilon\!+\!\omega)  
g_0^K({\bf x};\epsilon)\right].  
\label{Q}  
\end{equation}  
In equilibrium, $g^K_0= 2 \mbox{tanh}( \epsilon/2T)$,
and consequently
\begin{equation}  
Q_{eq}(\omega)=2 \hbar\omega \coth \frac{\hbar\omega}{2T}.  
\label{Qeq}  
\end{equation}  
Thus, $F_2$ vanishes, and $F_1$ reproduces the fluctuation--dissipation 
theorem: $F_1 = 2 Q_{eq} {\mathrm Re} {\cal D}_{\omega}({\bf x},{\bf  
x}';\omega)$.
 
Equations (\ref{3sum} -- \ref{D}) describe the noise in 
any system with a semiclassical potential. 
To obtain more explicit results, we have to specify a model.  
We assume that the average dwell time, $\tau_d$, is much larger than the  
classical ergodic time within the dot, and the contacts have  
the same number of channels.  Moreover, we consider only noise at frequencies  
$\omega\tau_d \ll 1$ not to be concerned with effects of  
dynamical screening.

The subsequent consideration follows the lines of  
the calculation of weak localization corrections\cite{Aleiner96}.  
The results for the factor (\ref{Q}) inside the dot  
(see Sec. IV of Ref.~\cite{Aleiner96}) take the form    
$Q=Q_{eq}+Q_{neq}$, where  
\begin{equation}  
Q_{neq}(\omega,V)=  
\frac{\sqrt{\Gamma}}{4}  
\sum_{\pm}  
\left[  
Q_{eq}(\omega\pm eV) -  Q_{eq}(\omega)   
\right],
\label{Qneq}  
\end{equation}
and $\Gamma$ is given by  (\ref{fff}). 
Within the same approximation, $Q$ in the leads is given by the 
equilibrium value (\ref{Qeq}).  
 
The appearance of the factor $\sqrt{\Gamma}$  
in (\ref{Qneq}) is not accidental. In the absence of   
the small angle scattering, $g_0^K({\bf x})$ can take only two possible  
values: either the equilibrium value of the left lead or that of the   
right lead. This is because a point ${\bf x}$ in phase space   
has a unique trajectory connecting it to incoming trajectories from the   
leads. Consequently, Eq.~(\ref{Q}) implies that $Q({\bf x})=Q_{eq}$.  
However, diffraction, modeled by small angle scattering,   
allows for two distinct classical trajectories to reach the same point.  
At such a point, $g_0^K({\bf x})$ is a linear combination of the  
equilibrium values of the leads, and therefore $Q({\bf x})$ assumes a   
nonequilibrium value. Thus, the factor $\sqrt{\Gamma}$ expresses the   
probability of two classical trajectories to become close to each other   
so that a transition from one to another due to diffraction is possible.  
  
Having the function $Q_{neq}$ in the form (\ref{Qneq}), the problem of  
using Eqs.~(\ref{3sum}) and (\ref{F}) for  the evaluation of the noise  
spectrum becomes equivalent to the evaluation of the weak localization  
correction (Sec.~VI of Ref.~\cite{Aleiner96}) up to the replacement  
of the Cooperon with Eq.~(\ref{Qneq}): ${\cal C}(1,\bar{1})   
\to 2\pi\nu Q_{neq}$. We, thus, obtain for the noise spectrum:   
\begin{equation}  
S(\omega) = G \left[Q_{eq}(\omega) + \frac{1}{2}\sqrt{\Gamma  
}Q_{neq}(\omega,V)  
\right],   
\label{S1}  
\end{equation}  
where $G$ is the conductance of the dot and $Q_{eq}$ is given by  
Eq.~(\ref{Qeq}). Substituting Eq.~(\ref{Qneq}) into Eq.~(\ref{S1}), we  
obtain the final result for the noise spectrum   
in the quantum dot with symmetric contacts:  
\begin{eqnarray}  
S(\omega)=  G \left\{  
Q_{eq}(\omega) +   
\frac{F}{2} \sum_{\pm} \left[Q_{eq}  
\left(\omega \pm \frac{eV}{\hbar}\right) - Q_{eq}\left(\omega\right)  
\right] \right\}, \nonumber  
\end{eqnarray}  
where $F$ is given by Eq.~(\ref{fff}).  
In the limit $T=0$,  $Q_{eq}(\omega)=2\hbar|\omega|$ and  
$S(\omega)$  at $\omega=0$ reduces to Eq.~(\ref{cccf}).   
  
To summarize, we constructed a theory for the quantum-to-classical crossover  
of the shot noise in mesoscopic system. A key ingredient of this crossover  
is the divergence of classical orbits in chaotic systems. This divergence is  
determined by the Lyapunov exponent. Thus, measurements of the  
of the shot noise in quantum dots as function of the dwell time can be used   
to determine the Lyapunov exponent of the underlying classical system.  
 
We would like to thank Carlo Beenakker for useful communications. 
O.A.~is grateful for discussions with David Menashe.  
This research was supported by Grant No.~9800065 from the USA-Israel   
Binational Science Foundation (BSF). I.A. is a Packard research fellow.  
  
\end{multicols}  

\begin{references}  
\vspace {-1.5cm}  
\bibitem{review}For review,   
M.~J.~M.~de Jong and C.~W.~J.~Beenakker in  
{\em Mesoscopic Electron Transport}, eds. L.~L.~Sohn, L.~P.~Kouwenhoven, and   
G.~ Sch\"{o}n, NATO ASI Series E, Vol. 345, p. 225 ( Kluwer    
Publishing, Dordrecht, 1997).    
\bibitem{Schottky18} W.~Schottky, Ann. Phys (Leipzig){ \bf 57}, 541 (1918).  
\bibitem{Beenakker92}  
C.~W.~J.~Beenakker and  M.~Buttiker, Phys. Rev. B {\bf 46}, 1889 (1992)  
\bibitem{1/3} K.E.~Nagaev, Phys. Lett. A {\bf 169}, 103 (1992);   
B.L.~Altshuler, L.S.~Levitov, A.Yu.~Yakovets, JETP Lett. {\bf 59}, 
857 (1994).  
\bibitem{1/4} H.~U.~Baranger and P.~Mello, Phys.~Rev.~Lett. {\bf 73}, 142  
(1994);  
R.~A.~Jalabert.~J.-L.~Pichard, and C.~W.~J.~Beenakker,   
Europhys.~Lett.~{\bf 27}, 255 (1994).  
\bibitem{Beenakker91} C.~W.~J.~Beenakker and H.~van Houten, 
Phys. Rev. B {\bf 43}, 12066 (1991).
\bibitem{Lesovik89} G.B.~Lesovik, JETP Lett.~{\bf 49}, 592 (1989).  
\bibitem{comment} Our definition of diffraction involves any
scattering angle different from that predicted by classical mechanics.
It includes (but not exhaustes) scattering from singular
points of potential energy, considered in 
J.B. Keller, J. Opt. Soc. Amer. {\bf 52}, 116 (1962).
\bibitem{Aleiner96} I.L.~Aleiner and A.I.~Larkin, Phys.~Rev.~B {\bf 54},  
14423   
(1996).   
\bibitem{Nazarov94} A derivation, similar in spirit, for the diffusive  
systems was done in Yu.~V.~Nazarov, Phys.~Rev.~Lett. {\bf 73}, 134 (1994).  
\bibitem{Keldysh64} L.~V.~Keldysh,   
Sov. Phys. JETP~{\bf 20}, 1018 (1964).  
\bibitem{Gutzwiller90} M.~C.~Gutzwiller, {\it Chaos in Classical and    
Quantum Mechanics} (Springer, N.Y., 1990).    
\bibitem{Whitney98} R.~S.~Whitney, I.~V.~Lerner IV, and R.~A.~Smith
Phys. Rev. B {\bf 58}, 10343 (1998).
\bibitem{AGD}  A.A.~Abrikosov, L.P.~Gorkov,  
I.E.~Dzyaloshinskii, {\it Methods of Quantum Field Theory in  
Statistical Physics}, (Prentice--Hall, Englewood Cliffs, NJ, 1963).  
\bibitem{Eilenberger}G.~Eilenberger, Z.~Physik, Bd. {\bf 214}, 195 (1968);  
A.~I.~Larkin and Y.~N.~Ovchinnikov, Sov.~Phys.~JETP {\bf 41}, 960 (1977). 
\bibitem{pre}I.L. Aleiner and A.I. Larkin, Phys. Rev. E {\bf 55}, 1243
(1997).
\end{references}
\end{document}